\documentclass[12pt]{iopart}
\usepackage{iopams}
\usepackage{epsf}
\usepackage{euscript}

\def\bbbq{{\mathchoice {\setbox0=\hbox{$\displaystyle\rm Q$}\hbox{\raise
0.15\ht0\hbox to0pt{\kern0.4\wd0\vrule height0.8\ht0\hss}\box0}}
{\setbox0=\hbox{$\textstyle\rm Q$}\hbox{\raise
0.15\ht0\hbox to0pt{\kern0.4\wd0\vrule height0.8\ht0\hss}\box0}}
{\setbox0=\hbox{$\scriptstyle\rm Q$}\hbox{\raise
0.15\ht0\hbox to0pt{\kern0.4\wd0\vrule height0.7\ht0\hss}\box0}}
{\setbox0=\hbox{$\scriptscriptstyle\rm Q$}\hbox{\raise
0.15\ht0\hbox to0pt{\kern0.4\wd0\vrule height0.7\ht0\hss}\box0}}}}
\def\bbbz{{\mathchoice {\hbox{$\sf\textstyle Z\kern-0.4em Z$}}
{\hbox{$\sf\textstyle Z\kern-0.4em Z$}}
{\hbox{$\sf\scriptstyle Z\kern-0.3em Z$}}
{\hbox{$\sf\scriptscriptstyle Z\kern-0.2em Z$}}}}

\begin{document}
\renewcommand{\baselinestretch}{1.5}
\title{ Quantum Graphs: A simple model for Chaotic Scattering}
\author{Tsampikos Kottos$^1${\footnote {corresponding author:
tsamp@chaos.gwdg.de}}
and Uzy Smilansky$^2$ \\
$^1$ Max-Planck-Institut f\"ur Str\"omungsforschung, 37073 G\"ottingen,
Germany,\\
$^2$ Department of  Physics of Complex Systems,
The Weizmann Institute of Science, 76100 Rehovot, Israel}
\date{\today }

\begin{abstract}
We connect quantum graphs with infinite leads, and turn them to scattering systems.
We show that they display all the features which characterize quantum scattering
systems with an underlying classical chaotic dynamics: typical poles, delay time and
conductance distributions, Ericson fluctuations, and when considered statistically,
the ensemble of scattering matrices reproduce quite well the predictions of 
appropriately defined Random Matrix ensembles. The underlying classical dynamics
can be defined, and it provides important parameters which are needed for the quantum
theory. In particular, we derive exact expressions for the scattering matrix, and an
exact trace formula for the density of resonances, in terms of classical orbits,
analogous to the semiclassical theory of chaotic scattering. We use this in order to
investigate the origin of the connection between Random Matrix Theory and the
underlying classical chaotic dynamics. Being an exact theory, and due to its relative
simplicity, it offers new insights into this problem which is at the fore-front of
the research in chaotic scattering and related fields.\\

\hspace {-0.5cm} submitted to J. Phys. A Special Issue -- Random Matrix Theory
\end{abstract}


\section{\bf Introduction}
\label{sec:introduction}

Quantum graphs of one-dimensional wires connected at nodes were introduced already
more than half a century ago to model physical systems. Depending on the envisaged
application the precise formulation of the models can be quite diverse and ranges
from solid-state applications to mathematical physics 
\cite{A85,FJK87,ML77,A81,CC88,NYO94,Ibook,MT01,Z98,V98,E95,R83}. Lately, quantum 
graphs attracted also the interest of the quantum chaos community because they can 
be viewed as typical and yet relatively simple examples for the large class of 
systems in which classically chaotic dynamics implies universal correlations in the 
semiclassical limit \cite{KS97,KS99,A99,BK99,SS00,T00,BSW02,BG00,K01}. Up to now we 
have only a limited understanding of the reasons for this universality, and quantum 
graph models provide a valuable opportunity for mathematically rigorous investigations 
of the phenomenon. In particular, for quantum graphs an exact trace formula exists 
\cite{R83,KS97,KS99} which is based on the periodic orbits of a mixing classical 
dynamical system. Moreover, it is possible to express two-point spectral correlation 
functions in terms of purely combinatorial problems \cite{BK99,SS00,T00,BSW02}.

By attaching infinite leads at the vertices, we get non-compact graphs, for which a 
scattering theory can be constructed \cite{KS00,BG01,TM01}. They display many of the 
feature that characterize scattering systems with an underlying chaotic classical 
dynamics \cite{M73,S89,GR89,RBUS90,EVP99}, and they are the subject of the present 
paper. The {\it quantum} scattering matrix for such problems can be written explicitly, 
together with a trace formula for the density of its resonances. These expressions 
are the analogues of the corresponding semiclassical approximations available in the 
theory of chaotic scattering \cite{M73,S89,GR89}, albeit here they are exact. With 
these tools we analyze the distribution of resonances, partial delay times, and the 
statistics of the fluctuating scattering amplitudes and cross sections. Moreover, 
we address issues like the statistical properties of the ensemble of the scattering 
matrices, and conductance distribution. Finally we analyze the effect of non-uniform  
connectivity on the statistical properties of the scattering matrix. A main part of 
our analysis will be focused on the comparison of the statistical properties of the 
above quantities with the Random Matrix Theory (RMT) predictions 
\cite{M75,MPS85,JSA92,HILSS92,BB94,FS96b,FS96,FS97,BFB97,GM98,MB99}.

The paper is structured in the following way. In section (\ref {sec:definitions}), 
the mathematical model is  introduced and the main definitions are given. Section 
(\ref{sec:s-matrix}) is devoted to the derivation of the  scattering matrix for
graphs. The trace formula for the  density of resonances of quantum graphs is
presented in section (\ref {sec:classical}) which includes also the analysis of 
the underlying classical  system. The next section, is dedicated to the analysis 
of various statistical  properties of the $S$-matrix. Our numerical data are
compared both with the predictions of RMT and with the semiclassical expectations. 
Finally, in section (\ref {sec:star}), we analyze the class of star-graphs for which 
several results can be analytically derived. Our conclusions are summarized in 
section (\ref {sec:conclusions}).


\section{\bf Quantum Graphs: Definitions}
\label{sec:definitions}

We start by considering a {\it compact} graph ${\cal G}$. It consists of $V$ {\it 
vertices} connected by $B$ {\it bonds}. The number of bonds which emanate from each 
vertex $i$ defines the valency $v_i$ of the corresponding vertex (for simplicity we 
will allow only a single bond between any two vertices). The graph is called $v-regular$ 
if all the vertices have the same valency $v$. The total number of bonds is $B={1\over 
2} \sum_{i=1} ^V v_{i}$. Associated to every graph is its connectivity matrix $C$. It 
is a square matrix of size $V$ whose matrix elements $C_{i,j}$ take the values $1$ if 
the vertices $i,j$ are connected with a bond, or $0$ otherwise. The bond connecting 
the vertices $i$ and $j$ is denoted by $b \equiv (i,j)$, and we use the convention 
that $i<j$. It will be sometimes convenient to use the ``time reversed" notation, where 
the first index is the larger, and $\hat b \equiv (j,i)$ with $j>i$. We shall also use 
the directed bonds representation, in which $b$ and $\hat b$  are distinguished as two 
directed bonds conjugated by time-reversal. We associate the  natural metric to the 
bonds, so that $x_{i,j}\  (x_{j,i})$  measures the distance from the vertex $i\  (j)$ 
along the bond.  The length of the bonds are denoted by $L_{b}$ and we shall henceforth 
assume that they are {\it rationally independent}. The mean length is defined by 
$\left<L \right>\equiv (1/B) \sum_{b=1}^B L_b$ and in  all numerical calculations 
bellow it will be taken to be $1$. In the {\it directed- bond} notation $L_{b} = 
L_{\hat b}$.

The {\it scattering} graph ${\tilde {\cal G}}$ is obtained by adding leads which
extend from  $M (\leq V)$ vertices to infinity.  For simplicity we connect at most
one lead to any vertex. The valency of these  vertices increases to ${\tilde v}_i
=v_i+1$. The $M$ leads are denoted by the index $i$ of the vertex to which they are
attached while $x_i$ now measures the distance from the vertex along the lead $i$.

The Schr\"odinger operator (with $\hbar=2m=1$) is defined on the graph ${\tilde {\cal
G}}$ in the following way: On the bonds $b$, the components $\Psi_b$ of the total wave
function $\Psi$ are solutions of the one - dimensional equation
\begin{equation}
\label{schrodinger}
\left( -i{\frac{{\rm d\ \ }}{{\rm d}x}}-A_b\right) ^2\Psi _b(x)=k^2\Psi
_b(x), \,\,\,\,\,\,\,\,\,\,\,\, \bigskip\ b=(i,j)
\end{equation}
where $A_b$ (with $\Re e(A_{b})\ne 0$ and $A_{b}= -A_{\hat b}$) is a ``magnetic
vector potential" which breaks the time reversal symmetry. In most applications
we shall assume that all the $A_{b}$'s are equal and the bond index will be dropped.
The components of the wave functions on the leads, $\Psi_i(x)$, are solutions of
\begin{equation}
\label{schrodinger1}
-{\frac{{\rm d^2\ \ }}{{\rm d}x^2}}\Psi _i(x)=k^2\Psi_i(x),
\,\,\,\,\,\,\,\,\,\,\,\, \bigskip\ i=1,...,M.
\end{equation}
At the vertices, the wavefunction satisfies boundary conditions which ensure current 
conservation. To implement the boundary conditions, the components of the wave function 
on each of the bonds $b$ and the leads $i$ are expressed in terms of counter propagating  
waves with a wave-vector $k$:
\begin {eqnarray}
&&{\rm On\ the \  bonds:\ } \Psi_{b} = a_{b} {\rm e}^{i(k+A_{b})x_{b}}
+c_{ b} {\rm e}^{i(-k+A_{b})x_{b}} \nonumber  \\
&&{\rm On\  the \ leads\ :\ } \Psi_i =  I_{i} {\rm e}^{-ikx_{i}} + O _{i}
{\rm e}^{ ikx_{i}} \ .
\label{wfun1}
\end{eqnarray}
The amplitudes $a_b,c_b$ on the bonds and $I_i,O_i$ on the lead are related  by
\begin{eqnarray}
\label{vertexcond}
{\hspace {-20mm}}
{\left (
\begin {array} {c}
   O_i  \\  a_{i,j_1}\\ \cdot  \\   a_{i,j_{v_i}}
  \end {array}
\right )} =
\Sigma^{(i)}
{\left ( \begin{array} {c}
   I_i  \\ c_{j_1,i} \\ \cdot  \\ c_{j_{v_i},i}
  \end {array} \right ) }, \quad\quad \quad\quad
\Sigma^{(i)}=
{\left (
\begin{array}{cccc}
\rho^{(i)}  & \tau^{(i)}_{j_1} & \cdot   & \tau^{(i)}_{j_{v_i}} \\
\tau^{(i)}_{j_1} & \tilde \sigma^{(i)}_{j_1,j_1} & \cdot  &
\tilde \sigma^{(i)}_{j_1,j_{v_i}} \\
\cdot  & \cdot &\cdot &\cdot \\
\tau^{(i)}_{j_{v_i}} & \tilde \sigma^{(i)}_{{j_{v_i}},j_1} &
\cdot & \tilde \sigma^{(i)}_{j_{v_i},j_{v_i}} \\
\end{array}
\right)}.
\end{eqnarray}
These equalities impose the  boundary conditions at the vertices. The vertex scattering 
matrices $\Sigma^{(i)}_{j,j'}$, are $\tilde v_i \times \tilde v_i$ unitary symmetric 
matrices, and $j,j'$ go over all the $v_i$ bonds and the lead which emanate from $i$. 
The unitarity of $\Sigma^{(i)}$   guarantees current conservation at each vertex.

On the right hand side of (\ref{vertexcond}), the vertex scattering matrix $\Sigma^{(i)}$
was written explicitly in terms of the vertex reflection amplitude $\rho^{(i)}$, the
lead-bond transmission amplitudes $\{ \tau^{(i)}_{j}\}$, and the $v_i\times v_i$ bond-bond
transition matrix $\tilde \sigma^{(i)}_{j,j'}$, which is {\it sub unitary} ($|\det \tilde
\sigma^{(i)}|<1 $), due to the coupling to the leads.

Graphs  for which there are no further requirements on the $\Sigma^{(i)}$ shall be 
referred  to as {\it  generic}. It is often convenient to compute the vertex scattering
matrix  from a requirement that the wave function is continuous and satisfies $Neumann$
boundary conditions at that vertex. These graphs shall be referred to as Neumann graphs, 
and the resulting $\tilde \Sigma^{(i)}$ matrices read:
\begin {equation}
\label{Neumann}
\tilde \sigma^{(i)} _{j,j'}= {2 \over {\tilde v}}  -\delta _{j,j'};\quad
\tau^{(i)}_j =   {2 \over {\tilde v}};\quad
\rho^{(i)} ={2 \over {\tilde v}}-1\ .
\end{equation}

Vertices which are not coupled to leads have $\rho ^{i}=1 ,\ \tau^{(i)}_j =0$, while
the bond-bond transition matrix $\tilde \sigma^{(i)}_{j.j'}={2 \over v}-\delta _{j,j'}$ 
is unitary.


\section{\bf The S-matrix for Quantum Graphs}
\label{sec:s-matrix}

It is  convenient to discuss first graphs with leads connected to all the vertices
$M=V$. The generalization to an arbitrarily number $M\leq V$ of leads (channels) is
straightforward and will be presented at the end of this section.

To derive the scattering matrix, we first write the  bond wave functions using the 
two  representations which are conjugated by   ``time reversal":
\begin{eqnarray}
\label{wfun2}
\Psi_{b}(x_b) &=& a_{b} {\rm e}^{i(k+A_{b})x_{b}}
+c_{ b} {\rm e}^{i(-k+A_{b})x_{b}}\  = \nonumber \\
\Psi _{\hat b}(x_{\hat b}) &=&a_{\hat b}{\rm e}^{i(k+A_{\hat b})x_{\hat b}}+
c_{\hat b}{\rm e}^  {i(-k+A_{\hat b})x_{\hat b}}\  = \nonumber  \\
&=&a_{\hat b}{\rm e}^  {i(-k-A_{\hat b})x_b}{\rm e}^  {i(k+A_{\hat b})L_b} +
c_{\hat b}{\rm e}^  {i(-k+A_{\hat b})L_b}{\rm e}^  {i(k-A_{\hat b})x_b} \ .
\end{eqnarray}
Hence,
\begin{equation}
c_b=a_{\hat b}{\rm e}^  {i(k+A_{\hat b})L_b},\,\,\,\,
a_b=c_{\hat b}{\rm e}^  {i (-k+A_{\hat b})L_b }.\label{wfun3}
\end{equation}
In other words, but for a phase factor, the outgoing wave from the vertex $i$ in the 
direction $j$ is identical to the incoming wave at $j$ coming from $i$.

Substituting $a_b$ from Eq.~(\ref{wfun3}) in Eq.~(\ref{vertexcond}), and solving for 
$c_{i,j}$ we get
\begin{eqnarray}
\label{wfun4}
c_{i,j'}&=&\sum_{r,s} \left({\bf 1}-\tilde S_B(k;A) \right
)^{-1}_{(i,r),(s,j)} D_{(s,j)}
\tau _s^{(j)} I_j \nonumber \\
O_i&=&\rho^{(i)} I_i + \sum_{j'}\tau_{j'}^{(i)} c_{ij'}
\end{eqnarray}
where ${\bf 1}$ is the $2B\times 2B$ unit matrix. Here, the ``bond scattering matrix'' 
${\tilde S}_b$ is a sub-unitary matrix in the $2B$ dimensional space of directed bonds 
which propagates the wavefunctions. It is defined as $\tilde S_B(k,A)=D(k;A)\tilde R $, 
with
\begin{eqnarray}
\label{DandT}
D_{ij,i^{\prime }j^{\prime }}(k,A) &=&\delta _{i,i^{\prime }}\delta
_{j,j^{\prime }}{\rm e}^{ikL_{ij}+iA_{i,j}L_{ij}} \\
\ \tilde R_{ji,nm} &=&\delta _{n,i}C_{j,i}C_{i,m}{\tilde \sigma}
_{ji,im}^{(i)}.
\nonumber
\end{eqnarray}
$D(k,A)$ is a  diagonal unitary matrix which depends only on the metric properties of 
the graph, and provides a phase which is due to free propagation  on the bonds. The 
sub-unitary matrix $\tilde R$ depends on the connectivity and on the bond-bond transition 
matrices ${\tilde \sigma}$. It  assigns a scattering amplitude for transitions between 
connected directed bonds. $\tilde R$ is sub-unitary, since
\begin{equation}
\label{subR}
|\det \tilde R| = \prod_{i=1}^V |\det \tilde \sigma ^{(i)}|<1.
\end{equation}

Replacing $c_{i,j'}$ in the second of Eqs.~(\ref{wfun4}) we get the following relation 
between the outgoing and incoming amplitudes $O_i$ and $I_j$ on the leads:
\begin{equation}
\label{inout}
O_i = \rho^{(i)} I_i + \sum_{j'j r s}\tau_{j'}^{(i)}
\left({\bf 1}-\tilde S_B(k;A) \right )^{-1}_{(i,r),(s,j)} D_{(s,j)} \tau
_s^{(j)} I_j
\end{equation}
Combining (\ref{inout}) for all leads $i=1,\dots ,V$, we obtain the unitary $V\times V$ 
scattering matrix $S^{(V)}$,
\begin{equation}
\label{scatmat}
S^{(V)}_{i,j}  = \delta_{i,j} \rho^{(i)}
  + \sum_{r,s} \tau^{(i)}_r
\left ({\bf 1}-\tilde S_B(k;A) \right )^{-1}_{(i,r),(s,j)} D_{(s,j)}
\tau _s^{(j)}.
\end{equation}
From Eq.~(\ref{scatmat}), we see that the scattering matrix may be decomposed into 
two parts $S^{(V)}(k)=S^{D}+S^{fl}(k)$ which are associated with two well separated 
time scales of the scattering process. $S^{D}(=\delta_{i,j} \rho^{i})$ is the prompt 
reflection at the entrance vertex and induces a ``direct" component. In general, it
varies very slowly with energy, and is system dependent. On the other hand the
``chaotic" component of the $S$ matrix, $S^{fl}(k)$, starts by a transmission from
the incoming lead $i$ to the bonds $(i,r)$ with transmission amplitudes 
$\tau^{(i)}_{r}$. The wave gains a phase ${\rm e}^{i(k+A_{b})L_{b}}$ for each bond it 
traverses and a scattering amplitude $\tilde \sigma^{(i)}_{r,s}$ at each vertex. This 
multiple scattering inside the interaction region becomes apparent when the expansion
\begin{equation}
\label{multi}
({\bf 1}-\tilde S_B(k;A))^{-1}= \sum_{n=0}^{\infty} \tilde S_B ^n(k;A)
\,\,\,\,\,\, 
\end{equation}
is substituted in    (\ref{scatmat}). Eventually the wave is transmitted
from the bond $(s,j)$
to the lead $j$ with an  amplitude $\tau _s^{(j)}$. Explicitly,
\begin{equation}
S^{(V)}_{i,j}  = \delta_{i,j} \rho^{(i)} +  \sum_{t \in {\cal
T}_{i\rightarrow j}}
{\cal B}_{t} {\rm e}^{i (k l_t +  \Theta_t)}
\label {sexplicit}
\end{equation}
where ${\cal T}_{i\rightarrow j}$ is the set of the trajectories on $\tilde
{\cal G}$ which lead from $i$ to $j$.  ${\cal B}_{t}$ is the amplitude
corresponding to a path $t$  whose length and directed length are $l_t=
\sum_{b\in t}L_b$ and $\Theta_t=\sum_{b\in t} L_bA_b$  respectively. 
Thus the
scattering amplitude $S^{(V)}_{i,j}$ is  a sum of a large number of partial
amplitudes, whose complex  interference brings about the typical irregular
fluctuations of $|S^{(V)}_{i,j}|^2$ as a function of $k$.

One of the basic concepts in the quantum theory of scattering are the 
resonances.
They represent long-lived intermediate states to which bound states of a 
closed
system are converted due to coupling to continua. On a formal level, 
resonances
show up as poles of the scattering matrix $S^{(M)}$ occurring at complex 
wave-numbers
$\kappa_n = k_n - \frac i2 \Gamma_n$, where $k_n$ and $\Gamma_n$ are the 
position
and the width of the resonances, respectively. From (\ref{scatmat}) it 
follows that
the resonances are the complex zeros of
\begin{equation}
\label {resonancecond}
\zeta_{\tilde {\cal G}}(\kappa) = \det \left({\bf 1} -\tilde S_B(\kappa
;A)\right)=0 \ .
\end{equation}
The eigenvalues of $\tilde S_B$ are in the unit circle, and therefore the
resonances appear in the lower half of the complex $\kappa$ plane. Moreover
from
Eq.~(\ref{resonancecond}) it is clear that their formation is closely
related to the
internal dynamics inside the scattering region which is governed by $\tilde
S_B$.

There exists an intimate link  between the scattering matrix and the
spectrum of the
corresponding closed graph. It manifests the exterior -interior duality
\cite{DS92a} for  graphs. The spectrum of the closed graph is the set of
wave-numbers for which
$S^{(V)}$ has $+1$ as an eigenvalue. This corresponds to a  solution where
no currents flow in
the leads so that the conservation of current is  satisfied on the internal
bonds.
$1$ is in the spectrum of
$S^{(V)}$ if
\begin{equation}
\label{phsca4}
\zeta_{\cal G}(k)=\det\left[{\bf 1} -S^{(V)}(k)\right] = 0 \quad .
\end{equation}
  Eq.~(\ref{phsca4}) can be transformed in an alternative  form
\begin{equation}
\label{phscaa} {\hspace {-15mm}}
\zeta_{\cal G}(k)= \det [ {\bf 1} -\rho ]
{\det[{\bf 1}-D(k)R] \over \det[{\bf 1}-D(k) {\tilde R}]} = 0 \quad ; \quad
R_{i,r;s,j} = \tilde R_{i,r;s,j} +\delta_{r,s}
{\tau^{(r)}_{i}\tau^{(r)}_{j}\over 1-\rho^{(r)} }.
\end{equation}
which is satisfied once
\begin{equation}
\label{phscaa1}
\det[{\bf 1}-D(k)R] = 0 .
\end{equation}
In contrast to ${\tilde R}$, $R$ is a unitary matrix in the space of
directed bonds, and
therefore the spectrum is real. (\ref {phscaa1}) is the secular equation
for the
spectrum of the compact part of the graph, and it was derived in a
different way in \cite
{KS97}.

The difference $\delta R = R - \tilde R$ gets smaller as larger graphs are considered 
(for graphs with Neumann  boundary conditions it is easy to see that the difference 
is of order $\frac{1}{v}$). That  is, the leads are weekly coupled to the compact part 
of the graph, and one can use  perturbation theory for the computation of the resonance 
parameters. To lowest order, $(\delta R =0)$, the resonances coincide with the spectrum 
of the compact graph. Let $k_n$ be in the spectrum.  Hence, there exists a vector 
$|n \rangle$ which satisfies the equation
\begin{equation}
D(k_n) R\  |n \rangle = 1\ |n \rangle \ .
\end{equation}
To first order in $\delta R$, the resonances acquire a width
\begin{equation}
\label{pertu3}
\delta \kappa_n = -i  {\langle n|D(k_n)\delta R|n\rangle \over \langle n|L|n\rangle} \,.
\end{equation}
To check the usefulness of this result, we searched numerically for the 
true poles
for a few scattering graphs and compared them with the approximation 
~(\ref{pertu3}).
In figure~1 we show the comparison for fully connected Neumann graphs
with $V=5,15$ and $A=0$.  As expected the agreement between the exact 
poles and the
perturbative results improves as $v$ increases.

\begin{figure}
\begin{center}
\epsfxsize.5\textwidth%
\epsfbox{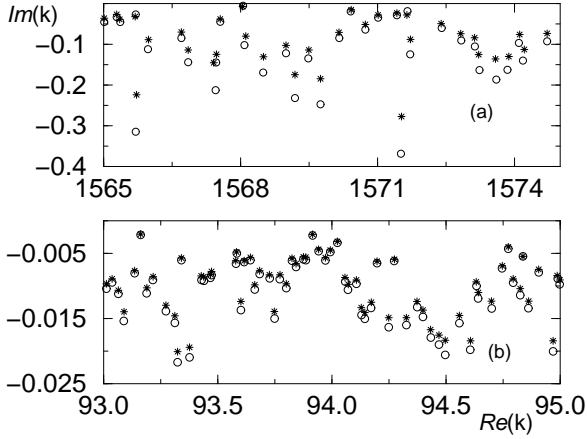}
\caption { Poles of the $S^{(V)}$-matrix for regular Neumann graphs. The exact 
evaluated poles are indicated with $(\circ )$ while $(\star )$ are the results 
of the perturbation theory (\ref{pertu3}): (a) $V=5$ and $v=4$ and (b) $V=15$ 
and $v=14$.}
\end{center}
\label{fig:fig2}
\end{figure}

We finally comment that the formalism above can be easily modified for graphs where
not all the vertices are attached to leads. If the vertex $l$ is not attached, one
has to set $\rho^{(l)}=1, \tau^{(l)}_j =0$  in the definition of $\Sigma^{(l)}$.
The dimension of the scattering matrix is then changed accordingly.

For the sake of completeness we quote here an alternative expression for the $S$ 
matrix which applies for Neumann graphs, exclusively  \cite{KS99}. For this purpose 
we define the $V\times V$ matrix
\begin{equation}
\label{secu1}
h_{i,j}(k,A) =\left\{
\begin{array}{cc}
-\sum_{m\neq i}C_{i,m}\cot(kL_{i,m})  , &
i=j \\ \\
C_{i,j}{\rm e}^  {-iA_{i,j}L_{i,j}}(\sin (kL_{i,j}))^{-1}, & i\neq j \
\end{array}
\right.   \nonumber
\end{equation}
in terms of which,
\begin{equation}
S^{(V)} = (i {\bf 1} + h(k))^{-1} (i {\bf 1}-h(k))
\label{phsca3}
\end{equation}
where ${\bf 1}$ is the $V\times V$ unit matrix. $S^{(V)}$ is unitary 
since $h(k)$ is
hermitian.

  In the case of graphs connected to leads at an arbitrary  set of $M<V$
vertices with indices
$\left \{i_l \right\}, \ \ l=1,\cdots,M$, the $M \times M$ scattering
matrix $S^{(M)}$
has to be modified in the following way
\begin{equation}
S^{(M)} = 2i W \left ( h(k) + i W^{T} W \right ) ^{-1} W^{T} -{\bf 1} \ .
\label{phsca5}
\end{equation}
Here $W_{i_l,j}=\delta_{i_l,j}$ is the $M\times V$ leads - vertices coupling
matrix, ( $W={\bf 1}$ when $M=V$). This form of the $S$ matrix is
reminiscent of the
expression  which was introduced by Weidenm\"uller to generalize the
Breit-Wigner theory
for  many channels and internal states. However, (\ref {phsca5}) is an
exact expression
which involves no truncations.

  It follows from (\ref {phsca5}) that in the present case, one can identify
the poles of the
$S$ matrix with the zeros of
\begin{equation}
\label{resonN}
\zeta_{\tilde {\cal G}}(\kappa) = \det \left(h(\kappa) + i W^{T} W
\right)=0 \ .
\end{equation}
Its main advantage over (\ref{resonancecond}) is that it involves a determinant of a 
matrix of much lower dimension.

\section{\bf The trace formula for resonances and  classical scattering on
graphs}
\label{sec:classical}

In the spectral theory of bounded hamiltonian systems, the most fundamental object of 
study is the spectral density which consists of a sum of $\delta$ functions at the 
spectral points. For open systems, it is replaced by the resonance density, which
is defined on the real $k$ line and consists of an infinite sum of Lorentzians which
are centered at $k_n=\Re e \kappa_n$ and have width $\Gamma _n =-2\Im m \kappa_n$,
where $\kappa_n$ are the complex poles of the scattering matrix. In the present section 
we shall express the resonance density for scattering graphs in terms of periodic 
orbits of their compact part. This is the analogue of the trace formula for bounded 
graphs.

\subsection{\bf The Trace Formula}

The resonance density $d_R(k)$ can be deduced from the total phase $\Phi(k) = \frac{1}
{i} \ln \det S^{(M)}(k) $ of the scattering matrix \cite {Krein}:
\begin{equation}
\label{resden}
d_R(k)\equiv{1\over 2\pi} {d\Phi\over dk} = -i{1\over 2\pi}{\partial\over
\partial k}\ln \det S^{(M)}(k).
\end{equation}
It is a smooth function for real $k$, and can be interpreted as the mean length of
the {\it delay} associated with the scattering at wave-number $k$ 
\cite{S89}.

Using Eq.~(\ref{scatmat}) and performing standard manipulations \cite{MW69}, we obtain 
the following expression for the phase $\Phi(k)$
\begin{equation}
\Phi(k)- \Phi(0) = -2 \Im m \ln \det (I- \tilde S_B(k;A)) + {\cal L}k \, .
\label{tra}
\end{equation}
where ${\cal L} =2\sum_{b=1}^B L_b $ is twice the total length of the 
bonds of
$\tilde {\cal G}$.

Using the expansion
\begin{equation}
\label{expan}
\ln \det (I- \tilde S_B(k;A)) = -\sum_{n=1}^{\infty} {1\over n} {\rm tr}
{\tilde
S_B}^n(k;A)
\end{equation}
we rewrite Eq.~(\ref{tra}) as
\begin{equation}
\label{tra1}
\Phi(k)= \Phi(0) + {\cal L}k +
2 \Im m \sum_{n=1}^{\infty} {1\over n} {\rm tr} {\tilde S_B}^n (k;A).
\end{equation}

On the other hand, using Eq.~(\ref{DandT}) we can write ${\rm tr} {\tilde
S_B}^n(k;A)$
as sums over $n-$periodic orbits on the graph
\begin{equation}
\label{ppo}
{\rm tr} ({\tilde S_B}^n(k;A)) = \sum_{p\in {\cal P}_{n}} n_p {\tilde {\cal
A}_p}^r
{\rm e}^  {i(l_p k+\Theta_p)r} \quad ,
\end{equation}
where the sum is over the set ${\cal P}_n$ of primitive periodic orbits 
whose
period $n_p$ is a divisor of $n$, with $r=n/n_p$ (primitive periodic orbits
are those which
cannot be written as a repetition of a shorter periodic orbit). The 
amplitudes
${\tilde {\cal A}}_p$ are the products of the bond-bond scattering
amplitudes $\tilde
\sigma^{(i)}_{b,b'}$ along the primitive loops i.e.
\begin{equation}
\label{ampli}
{\tilde {\cal A}_p} = \prod_{i=1}^{n_p} {\tilde \sigma^{(i)}_{b,b'}} \ .
\end{equation}
Substituting Eqs.~(\ref{tra1},\ref{ppo}) in Eq.~(\ref{resden}) one gets the resonance
density
\begin{equation}
\label{resdens}
d_R(k)  = {1\over 2 \pi}{\cal L}+{1\over \pi}\Re e\sum_{n=1}^{\infty}\sum_{p\in 
{\cal P}_{n}} n_p l_p r  {\tilde {\cal A}_p}^r {\rm e}^  {i(l_p k+\Theta_p)r}
\end{equation}
Eq.~(\ref{resdens}) is an {\it exact} trace formula for the resonance density. The
first term on the right hand side of Eq.~(\ref{resdens}) corresponds to the smooth
resonance density, while the second provides the fluctuating part. We notice that
the mean resonance spacing is given by
\begin{equation}
\label{mrd}
\Delta= {2\pi\over {\cal L}} \simeq {2 \pi\over 2 B \langle L\rangle}
\end{equation}
and it is the same as the mean level spacing obtained for the corresponding bounded 
graph ${\mathcal G }$ \cite{KS97}.

\subsection{\bf Classical Dynamics}

We conclude this section with the discussion of the classical dynamics on the graph
${\tilde {\cal G}}$. A classical particle moves freely as long as it is on a bond.
The vertices are singular points, and it is not possible to write down the analogue
of the Newton equations there.  In \cite {KS97,KS99} it was shown that it is possible
to define a classical evolution on the graph: A {\it Poincar\'{e} section} on the
graph consists of the discrete set of directed bonds. The phase-space density at a
(topological) time $n$ is the set of occupation probabilities $\rho_b(n)$ of the
directed bonds, and the classical evolution is governed by a Markovian master equation.
Applied to the compact part of a scattering graph it reads,
\begin{equation}
\label{master}
\tilde \rho _b(n+1)=\sum_{b^{\prime }} \tilde  U_{b,b^{\prime }}
\tilde \rho _{b^{\prime }}(n)
\end{equation}
where the transition matrix $ \tilde U_{b,b^{\prime }}$ is given by the 
corresponding
{\it quantum} transition probability
\begin{equation}
\tilde U_{ij,nm}= \left|\tilde R_{ij,nm}\right|^2=\delta _{j,n} |
\sigma^{(j)}_{ij,jm}|^2.
  \label{cl3}
\end{equation}
  Notice that $\tilde U$ does not involve any metric information on the 
graph.

  Due to loss of flux to the leads $\sum_{b'}{\tilde U}_{bb'}<1$, and the
phase-space measure is not preserved, but rather, decays in time. The
probability to remain on $\tilde {\cal G}$ is
\begin{equation}
{\tilde P}(n) \equiv \sum_{b=1}^{2B}{\rho}_b(n) = \sum_{b,b'}{\tilde
U}_{bb'}{\rho}_{b'}(n-1)
\simeq  {\rm e}^  {-\Gamma_{cl}n}{\tilde P(0)}
\end{equation}
where $\exp (-\Gamma_{cl})$ is the largest eigenvalue of the ``leaky" 
evolution
operator ${\tilde U}_{bb'}$.

For the $v$-regular graph with $\tilde \sigma^{(i)}$ given by (\ref 
{Neumann})
the spectrum of $\tilde U$ is restricted to the interior of a circle with
radius given by the maximum eigenvalue $\nu_1=(v-1)\tau^2+\rho^2$ with
corresponding eigenvector $|1>=(1/2B)(1,1,\cdots,1)^T$. Hence the decay rate
$\Gamma_{cl}=-\ln \nu_1$ for regular Neumann graphs take the simple form
\begin{equation}
\label{Ndrate}
\Gamma_{cl}=-\ln \left(1-\tau^2\right) \approx (2/(1+v))^2.
\end{equation}
We notice that removing the leads from the vertices and turning ${\tilde
{\cal G}}$ into a compact graph ${\cal G}$ we get $\Gamma_{cl}= 0$ since in
this case $\tau^{(i)}=0$ (and $\rho^{(i)}=1$) and the phase-space measure
is preserved as expected.

The inverse decay rate $T_{cl} = \Gamma_{cl}^{-1}$, gives the average classical delay 
time that the particle spends within the interaction region. Injecting a particle from 
the leads to the scattering domain, its probability to be on any bond randomizes, 
because at each vertex a Markovian choice of one out of $v$ directions is made. The 
longer a particle remains within the interaction regime, the more scattering events it 
experiences. The set of trapped trajectories whose occupancy decays exponentially in 
time is the analogue of the strange repeller in generic Hamiltonian systems displaying 
``chaotic scattering".


\section{\bf Statistical Analysis of the $S-$matrix}
\label{sec:statistics}


So far we developed the scattering theory of graphs, pointing out their similarity
with scattering systems which display chaotic scattering in the classical limit. 
Due to the interference of a large number of amplitudes, the $S-$matrix fluctuates 
as a function of $k$, and its further analysis calls for a statistical approach 
which will be the subject of the present section. We shall show that quantum graphs 
possess typical poles, delay time and conductance distributions, Ericson fluctuations 
of the scattering amplitudes, and when considered statistically, the ensemble of 
scattering matrices are very well reproduced by the predictions of RMT. At the same 
time deviations from the universal RMT results, which are related to the system-
specific properties of some graphs, will be  pointed out. The study  of these 
deviations is especially convenient for graphs because of the transparent and simple 
scattering theory developed in terms of scattering trajectories.

An important parameter which is associated with the statistical properties of the 
$S$-matrix is the Ericson parameter defined through the scaled mean resonance width 
as:
\begin{equation}
\label {eripar}
\left<\gamma\right>_k \equiv  {\left<\Gamma_n\right>_k  \over \Delta}
\end{equation}
where $\langle \cdot \rangle _k$ denotes spectral averaging and $\Delta$ is the
mean spacing between resonances. The Ericson parameter determines whether the
resonances overlap $(\langle \gamma\rangle_k >1)$ or are isolated $(\langle
\gamma\rangle_k <1)$. Typical example for the two extreme situations are shown
in figure~2.

\begin{figure}
\begin{center}
\epsfxsize.5\textwidth%
\epsfbox{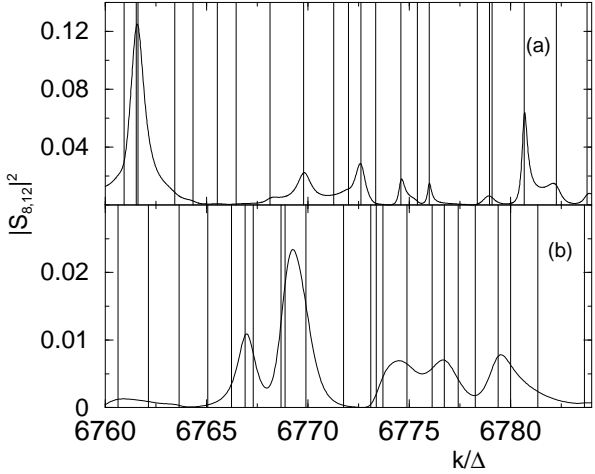}
\end{center}
\caption{ Representative examples of scattering cross sections 
$|S_{i,j}^{(V)}|^2$
for Neumann graphs in a regime of  (a) isolated resonances ($V=15$, 
$v=14$ with
corresponding $\left<\gamma\right>_k=0.59$) and (b) overlapping 
resonances ($V=49$,
$v=14$ with corresponding $\left<\gamma\right>_k\simeq 2$). The real 
parts of the
resonances in this energy interval are indicated by the vertical lines.}
\end{figure}

The degree of resonance overlap determines the statistical properties of the
$S$-matrix.  We shall show in the sequel that the mean width  can be
approximated  by the classical decay rate $\langle \gamma\rangle_k=
\gamma_{cl}$. For the $v$
regular graphs discussed above, we have
\begin{equation}
\label{clmr}
\gamma_{cl}\equiv {\Gamma_{cl}\over \Delta} \approx {4\over 2\pi} 
{v\over 1+v}
{V\over 1+v}.
\end{equation}
where we made use of Eqs.~(\ref{mrd},\ref{Ndrate}).
Thus changing $v$ and $V$ we can control the degree of overlap allowing to
test various phenomena.

In what follows, unless explicitly specified, we shall consider regular
graphs with
one lead attached to each vertex, i.e. $M=V$. Finally, the widths are
always scaled
by the mean spacing $\Delta$ i.e. $\gamma_n\equiv {\Gamma_n \over \Delta}$.

\subsection{\bf The resonance width distribution }

The resonance width distribution can be computed for a given graph in the following 
way. Consider the complex $\kappa=x+iy$ plane, where the zeros of the secular function
\begin{equation}
\label {resonancecond1}
\zeta_{\tilde {\cal G}}(\kappa) = \det \left({\bf 1} -\tilde S_B(\kappa ;A)\right) = f(x,y) +i
g(x,y)
\ .
\end{equation}
are the poles of the $S$ matrix (resonances). The variables $(x,y)$
are expressed in units of $\Delta$. On average, the number of resonances
with real part, $x \in
[0,X]$ is $X$. The density of resonance widths reads
\begin{equation}
\hspace{-10mm}
{\mathcal P}(\gamma) =\lim_{ X \rightarrow \infty} \frac{1}{ X } \int_0^X
\delta\left (f(x,y= - \gamma)\right ) \delta\left (g(x,y= -  \gamma)\right )
\left |\frac{{\rm d}\zeta_{\tilde {\cal G}}(\kappa)}{ {\rm d} \kappa}\right |^2_{y= -
\gamma}   {\rm d}x
\ .
\label{eq:disty}
\end{equation}
We used the Cauchy-Riemann theorem for the evaluation of the Jacobian $(f_xg_y - f_y 
g_x)$ which multiplies the two $\delta$ functions that locate the complex zeros of 
$\zeta_{\tilde {\cal G}}$. We now recall that the $\kappa$ dependence of $\tilde S_B
(\kappa ;A)$ comes from the factors ${\rm e}^{i\kappa L_b}$ in (\ref {DandT}). This
implies that for a given $y$, $\zeta_{\tilde {\cal G}}$ is a quasi periodic function 
of $x$. Moreover, expanding the determinant (\ref {resonancecond1}) it is not difficult 
to show that the frequencies involved can be written as linear combinations of the bond 
lengths $\lambda =\sum  q_b L_b $ with integer coefficients $q_b =0, 1, {\rm or}\ 2$. 
Since the bond lengths are rationally independent, we find that $\zeta_{\tilde {\cal G}}$
depends on a {\it finite}  number of incommensurate frequencies. The expression (\ref
{eq:disty}) can be regarded as a ``time"  integral over a trajectory on a multidimensional
incommensurate torus, which covers the torus ergodically. Hence, the integral can be
replaced by a phase space average,

\begin{equation}
\hspace{-10mm}
{\mathcal P}(\gamma) =
\int_0^{2\pi}\frac {{\rm d} \psi_1}{2\pi} \cdots \int_0^{2\pi}  \frac 
{{\rm d}
\psi_J}{2\pi} \
\delta\left (f({\vec \psi} , - \gamma)\right )\ \delta\left (g({\vec \psi} ,
-\gamma)\right ) \
\left |\frac{{\rm d}\zeta_{\tilde {\cal G}}({\vec \psi,-\gamma})}{ {\rm d}
\kappa}\right |^2_ .
\label{eq:distyergod}
\end{equation}
Here ${\vec \psi}$ denotes  the vector of independent angles on the $J$ dimensional
torus. Although the above formula provides a general framework, its application to
actual graphs is a formidable task.

An important feature of the distribution of the resonances in the complex plane can
be deduced by studying the secular function $\zeta_{\tilde {\cal G}}(\kappa)$. Consider
$\zeta_{\tilde {\cal G}}(\kappa=0)$. If one of the eigenvalues of the matrix $\tilde R$
(\ref {DandT}) takes the value $1$, $\zeta_{\tilde {\cal G}}(k=0) =0$ and because of
the quasi-periodicity of $\zeta_{\tilde {\cal G}}$, its zeros reach any vicinity of
the real axis infinitely many times. The largest eigenvalue of the $\tilde R$ matrix
for $v$-regular Neumann graphs is $1$, and therefore the distribution of resonance 
widths is  finite in the vicinity of $\gamma=0$. For generic graphs, the spectrum of 
$\tilde R$ is inside  a circle of radius $\lambda_{max} <1$. This implies that the 
poles are excluded from a strip just under the real axis, whose width can be estimated 
by
\begin{equation}
\label{gapKS}
\Gamma_{gap}=-2 \ln( |\lambda_{max}|)/L_{max}.
\end{equation}
where $L_{max}$ is the  maximum bond length. The existence of a gap is 
an important
feature of the resonance width distribution ${\cal P}(\gamma)$ for 
chaotic scattering
systems.

A similar argument was used recently in \cite{BG01} in order to obtain an {\it upper}
bound for the resonance widths.  It is
\begin{equation}
\label{upl}
\Gamma_{max}\sim -2 \ln( |\lambda_{min}|)/L_{min}\ ,
\end{equation}
where $\lambda_{min}$ and $L_{min}$ are the minimum eigenvalue and bond length,
respectively.

The distribution of the complex poles  for a generic fully connected graph with $V=5$
is shown in figure~3a. The vertical line which marks the region from which resonances
are excluded was computed using (\ref {gapKS}).
\begin{figure}
\begin{center}
\epsfxsize.5\textwidth%
\epsfbox{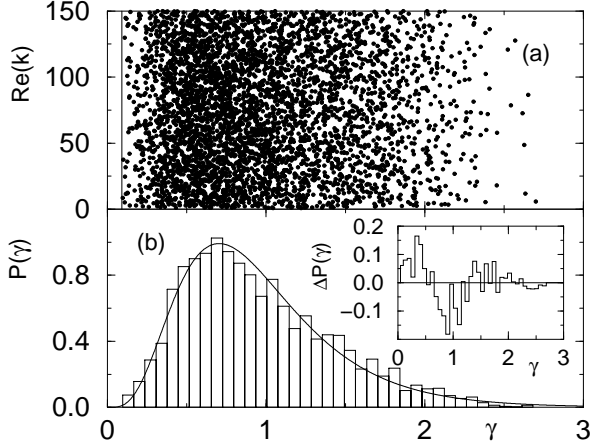}
\caption{ (a) The 5000 resonances of a single realization of a complete $V=5$ graph 
with $A\ne 0$ and $M=V$. The solid line marks the position of the gap $\gamma_{\it gap}$. 
(b) The distribution of resonance widths ${\cal P}(\gamma)$. The solid line is the RMT 
prediction (\ref{FSpole}). The difference ${\cal P}(\gamma)-{\cal P}_{CUE} (\gamma)$ 
is shown in the inset.}
\end{center}
\end{figure}

Random matrix theory can provide an expression for the distribution of resonances. In the
case of non-overlapping resonances,  perturbation theory  shows that the resonance widths 
are distributed according to the so-called $\chi^2$ distribution
\begin{equation}
\label{chi2}
{\cal P}(\gamma) = \frac {(\beta M/2)^{\beta M/2}}{\langle \gamma\rangle_k{\it \Gamma}
(\beta M/2)} \left({\gamma\over\langle \gamma\rangle_k}\right)^{\beta M/2 -1} 
\exp(-\gamma \beta M/2\langle \gamma\rangle_k ),
\end{equation}
where $\beta=1(2)$ for systems which respect (break) time-reversal symmetry, and 
${\it \Gamma}(x)$ is the gamma-function. Once $\langle \gamma\rangle_k$ becomes large 
enough the resonances start to  overlap, and (\ref{chi2}) does not hold. In the general 
case, Fyodorov and Sommers \cite{FS96,FS97} proved that the distribution of scaled 
resonance widths for the unitary random matrix ensemble, is given by
\begin{equation}
\label{FSpole}
{\cal P}(\gamma) = \frac {(-1)^M}{\Gamma(M)} \gamma^{M-1} {d^M\over 
d\gamma^M}
\left({\rm e}^  {-\gamma \pi g} {\sinh(\gamma\pi)\over(\gamma\pi)}\right)
\end{equation}
where the parameter $g={2\over (1-\langle S^D\rangle_k^2)} -1$ controls the degree of
coupling with the channels (and is assumed that $g_i=g \,\,\forall i=1,...M$). For
$g\gg 1$ (i.e. weak coupling regime) Eq.~(\ref{FSpole}) reduces to
(\ref{chi2}).

In the limit of $M\gg 1 $, Eq.~(\ref{FSpole}) reduces to the following expression
\cite{FS97}
\begin{equation}
{\cal P}(\gamma)=
\left\{
\begin{array}{lll}
{M\over 2\pi \gamma^2}&\ {\rm for}\ &{M\over \pi(g+1)}<
\gamma<{M\over\pi(g-1)} \\\nonumber
\ 0&\ \ \ \ & \ \ \ \ \ \  {\rm otherwise} \nonumber
\end{array}
\right.  \ .
\label{largeM}
\end{equation}
It shows that in the limit of large number of channels there exist a  strip in the 
complex $\kappa-$ plane which is free of resonances. This is in agreement with 
previous findings \cite{GR89,M75,SZ88}. In the case of maximal coupling i.e. $g=1$, 
the power law (\ref{largeM}) extends to  infinity, leading to divergencies of the 
various moments of $\gamma$'s. Using  (\ref{FSpole}) we recover the well known 
Moldauer-Simonius relation \cite{M75} for the mean resonance width \cite{FS97}
\begin{equation}
\label{SMrw}
\langle \gamma\rangle _k = -\frac {\sum_{i=1}^{V} \ln (|\langle S^{D}\rangle _k|^2)}{2\pi}.
\end{equation}

The resonance width distribution for a $V=5$ regular and generic graph is shown in
figure~3b together with the RMT prediction, which reproduces the numerical distribution
quite well. Figure~4 shows a similar comparison for a Neumann graph. The relatively
high abundance of resonances in the vicinity of the real axis conforms with the
expectations.

\begin{figure}
\begin{center}
\epsfxsize.5\textwidth%
\epsfbox{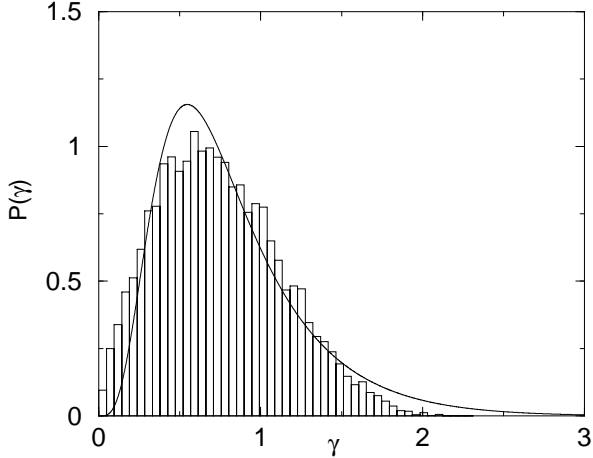}
\end{center}
\label{figure:n-resdist}
\caption {  Resonance width distribution ${\cal P}(\gamma)$ for a complete
Neumann graph with $M=V=5$ and $A\ne 0$. The solid line is the RMT prediction
(\ref{FSpole}). }
\end{figure}


\subsection{\bf The Form Factor}

To investigate further the dynamical origin of the resonance fluctuations, we study
the resonance two-point form factor $K(t)$. The main advantage of $K(t)$ is that it
allows us to study the resonance fluctuations in terms of classical orbits. It is
defined as
\begin{equation}
\label{rescor}
K_R(t)\equiv  \int {\rm d}\chi \ {\rm e}^{i 2\pi \chi {\cal L} t}
R_2(\chi).
\end{equation}
where $t$ measures lengths in units of the Heisenberg length $l_H={\cal L}$
and $
R_2(\chi)$ is the excess probability density of finding two resonances at a
distance
$\chi$,
\begin{equation}
\label{corr1}
\hspace {-15mm}
R_2(\chi;k_0) =\langle \tilde d_R  (k+ {\chi \over 2})\ \tilde d_R(k-
{\chi\over 2})\rangle_k = \langle {\Delta\over 2k_0}\int_{-k_0} ^{k_0}
{\tilde d}_R  (k+ {\chi \over 2})\ {\tilde d}_R(k- {\chi\over 2})
dk\rangle_{k_0}
\ .
\end{equation}
Above $\langle \ldots \rangle _{k_0}$ indicate averaging oven a number of spectral 
intervals of size $\Delta_k=2k_0$, centered around $k_0$. Here $\tilde d_R(k)$ is 
the oscillatory part of $d_R(k)$ (see Eq.~(\ref{resdens})). Substituting the latter 
in Eqs.~(\ref {rescor},\ref{corr1}) we obtain $K_R(t)$ in terms of periodic orbits 
and their repetitions
\begin{equation}
\label{ff1}
K_R(t)= {2{\cal N}\over{\cal L}^2}\left|\sum_p\sum_r l_p {\tilde {\cal 
A}}_p^r
{\rm e}^{i (kl_a+ \Theta_a)} \delta_{\cal N} (t-rl_p/{\cal L})\right|^2 .
\end{equation}
where $\delta_{\cal N}(x) = (\sin({\cal N}x/2))/({\cal N}x/2)$ and ${\cal N}={\Delta_k
\over\Delta}$. A similar sum contributes to the spectral form factor of the compact
graph \cite{KS97,KS99}. However, the corresponding amplitudes are different
due to the fact that ${\tilde {\cal A}}_p$ contains also the information about the
escape of flux to the leads.

\begin{figure}
\begin{center}
\epsfxsize.5\textwidth%
\epsfbox{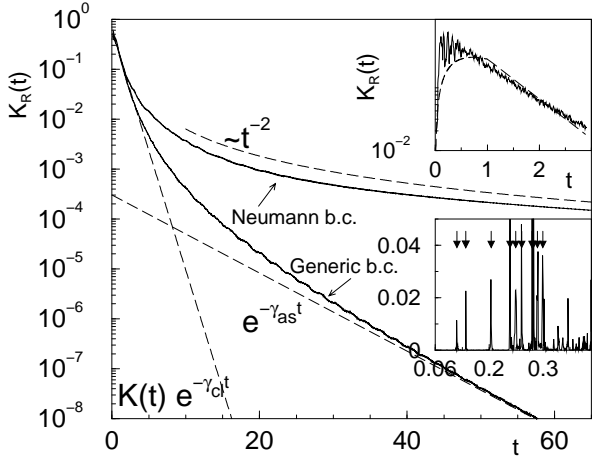}
\caption{The form factors $K_R(t)$ for a complete $V=5$ graph, with either generic or
Neumann  boundary conditions, $A\neq 0$ and $M=V$. The data were averaged over $5000$ 
spectral intervals and smoothed on small $t$ intervals. Upper inset: $K_R(t)$ for small 
times.  Solid line correspond to the numerical data for the pentagon with generic boundary 
conditions while dashed line is the approximant $K_R(t)\approx K(t) exp(-\gamma_{cl} 
t)$. Lower inset: $K_R(t)$ calculated with high resolution. The lengths of  periodic 
orbits of the close graph are indicated by arrows.}
\end{center}
\end{figure}

Assuming that all periodic orbits decay at the same rate, one would 
substitute
${\tilde {\cal A}}_p$ with ${\cal A}_p \exp(-n_p \gamma_{cl}/2)$ where 
${\cal
A}_p$ is the weight assign to the periodic orbit of the corresponding 
close system.
Then Eq.~(\ref{ff1}) takes the following simple form
\begin{equation}
\label{corr6}
K_R(t) \approx K(t) {\rm e}^  {-\gamma_{cl}t}\,\,\,,
\end{equation}
where $K(t)$ is the form factor of the compact system \cite{EVP99}. 
Notice that
for $t\ll \gamma_{cl}^{-1}$ the resonance form factor $K_R(t)$ is equal 
to $K(t)$.
This is  reasonable since an open system cannot be distinguished from a 
closed
one during short times. This simple approximation is checked in the 
inset of
figure~5 (see dashed line) and it is shown to reproduce the numerical 
data rather
well in the domain $t\le 5$. The asymptotic decay is dominated by the 
resonances
which are nearest to the gap, and it cannot be captured by the crude 
argument
presented above. For generic graph, $K_R$ decays exponentially but with 
a rate
given by $\gamma_{as} =\gamma_{gap}$ (the best fit, indicated in 
figure~5 by the
dashed line, give $\gamma_{as}$ which agrees with  $\gamma_{gap}$ within 
$30\%$).
For the graph with Neumann boundary conditions, $\gamma_{gap}=0$ and one 
expects
an asymptotic power-law decay. The corresponding best fit (see dashed 
line in
figure~5) shows $t^{-2}$.

In Hamiltonian systems in more than one dimension, the size of the 
spectral interval
$\Delta_k$ where the spectral average is performed is limited by the 
requirement
that the smooth part of the spectral density is approximately constant. 
Here instead
we can take arbitrarily large spectral intervals since the smooth 
spectral density
is constant \cite{KS97,KS99}. This way, one can reach the domain where 
the function
$K_R(t)$ is composed of arbitrarily sharp spikes which resolve 
completely the length
spectrum for lengths which are both smaller and larger than $l_H$. In 
the inset of
figure~5 we show the numerical $K(t)$ calculated with high enough 
resolution. In the
same inset we mark with arrows the location of the lengths of short 
periodic orbits.
We notice that as long as $t{\cal L}$ is shorter than the length of the 
shortest
periodic orbit, $K(t)=0$. With increasing $t$, the periodic orbits 
become exponentially
dense and therefore the peaks start to overlap, giving rise to a quasi 
continuum
described approximately by Eq.~(\ref{corr6}).

\subsection{\bf Ericson fluctuations}

As was mentioned above, the scattering cross sections are dominated by 
either isolated
resonances, or by overlapping resonances whose fluctuations follow a 
typical pattern.
These patterns were first discussed by Ericson \cite {E60} in the frame 
of nuclear
physics and were shown to be one of the main attributes of chaotic 
scattering \cite
{S89}. The transition between the two regimes is controlled by the 
Ericson parameter
(\ref {eripar}). Typical fluctuating cross sections are shown in figure~2.

\begin{figure}
\begin{center}
\epsfxsize.5\textwidth%
\epsfbox{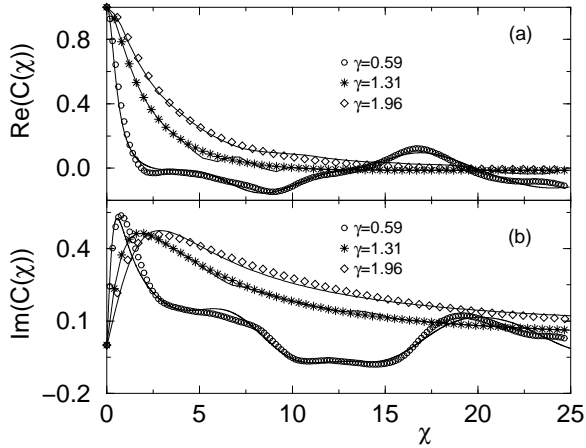}
\end{center}
\label{figu:autocorr}
\caption{ The autocorrelation function $C(\chi,\nu=1)$ for regular 
graphs with
Neumann boundary conditions. ($\circ$) correspond to a graph with
$\gamma=0.59$
(isolated resonance regime), ($\star$) to a graph with $\gamma=1.36$
(intermediate
regime) while ($\diamond$) to a graph with $\gamma \simeq 2$ (overlapping
resonances
regime). The solid lines correspond to the theoretical expression
(\ref{ericscl}):
(a) The real part of $C(\chi,\nu=1)$ ; (b) The imaginary part of
$C(\chi,\nu=1)$.}
\end{figure}

A convenient measure for Ericson fluctuations is the autocorrelation 
function
\begin{equation}
\label{ericcor}
C(\chi; \nu) =  {1\over\Delta j} \sum _{ j=j_{min}}^{j_{max}}\left 
\langle  \
S^{(M)}_{j,j+\nu}(k+{\chi\over 2}) \ S^{(M)\star}_{j,j+\nu}
(k-{\chi \over 2}) \ \right  \rangle_{k}
\end{equation}
where $\Delta j = j_{max}-j_{min}+1$. To evaluate Eq.~(\ref{ericcor}) we
substitute
the expression of the $S$-matrix from Eq.~(\ref{sexplicit}) and  split the
sum over
trajectories into two distinct parts: the contributions of short
trajectories are
computed explicitly by following the multiple scattering expansion up to
trajectories
of length $l_{max}$. The contribution of longer orbits are approximated by
using the
diagonal approximation, which results in a Lorentzian with a width
$\gamma_{Er}$.
Including explicitly up to $n=3$ scattering events we get,
\begin{eqnarray}
\label{ericscl}
C(\chi;\nu ) &\approx & G {\rm e}^  {il_{max}\chi} \frac {
\gamma_{Er}}{\gamma_{Er}-i\chi}
+\frac1{\Delta j}\sum_{j=j_{min}}^{j_{max}} [ \tau^4 {\rm e}^  {i\chi
L_{j,j+\nu}}
+\tau^4 \rho^4 {\rm e}^  {3i\chi L_{j,j+\nu} } \nonumber \\
&+& \tau^6 \sum_{m\neq j,j+\nu} {\rm e}^  {i\chi(L_{j,m}+L_{m,j+\nu})} ]
\end{eqnarray}
where $G$ is determined by the condition $C(\chi=0;\nu)=1$. The 
interplay between the
contributions of long and short periodic orbits is shown in figure~6. 
For overlapping
resonances, the autocorrelation function is well reproduced by a 
Lorentzian while for
the case of isolated resonances one can clearly see the contributions of 
short paths.

\begin{figure}
\begin{center}
\epsfxsize.5\textwidth%
\epsfbox{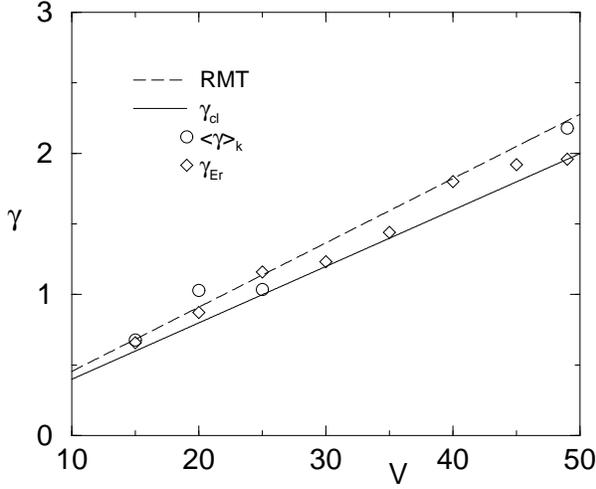}
\caption{ The mean resonance width $\left<\gamma\right>_k$, autocorrelation
width $\gamma_{\it Er}$, the classical expectation $\gamma_{cl}$, and 
the RMT
prediction (\ref{SMrw}) vs. $V$ for various graphs with Neumann boundary
conditions and constant valency $v=14$.}
\end{center}
\end{figure}

In figure~7 we report the mean resonance width $\langle \gamma\rangle_k$ calculated
numerically for  various graphs, the parameter $\gamma_{\rm Er}$ extracted from the
best fit of the  numerical $C(\chi)$ with Eq.~(\ref{ericscl}), together with the RMT
prediction Eq.~(\ref{SMrw}), and the classical expectation given by Eq.~(\ref{clmr}).
The results justify the use  of the classical estimate especially in the limit $V
\rightarrow \infty$ for fixed $v/V$ (which is the analogue of the semiclassical limit).
In this limit, the RMT and the classical estimate coincide.


\subsection{\bf $S-$matrix statistics}

One can check further the applicability of RMT by studying the entire $S^{(M)}-$matrix 
distribution function. The probability density of $M\times M$ unitary $S^{(M)}$-matrices 
is defined in a $M+\beta M(M-1)/2$ parameter space and was found first in \cite{MPS85} 
to be given by the Poisson Kernel
\begin{equation}
\label{sRMT}
d{\cal P}_{\bar S}(S^{(M)})=p_{\bar S}(S)
d\mu(S)=C_{\beta}\frac{[\det(1-{\bar S}
{\bar S^{\dagger}})]^{(\beta M +2-\beta)/2}}{|\det(1-{\bar S}{\bar
S^{\dagger}})|^{(\beta M +2-\beta)}} d\mu_{\beta}(S)
\end{equation}
where $C_{\beta}$ is a normalization constant which depend on the symmetry class. All
system-specific relevant informations are included in the ensemble average $S^{(M)}-$
matrix, defined as ${\bar S}_{i,j}= \langle S^D\rangle _k$. For regular graphs, ${\bar 
S}$ is proportional to the unit matrix i.e. ${\bar S}_{i,j}= \frac{1-v}{1+v}\delta_{i,j}$,
while the eigenvectors are distributed uniformly and independent from the eigenphases.
The invariant measure $d\mu_{\beta}(S)$ is given as
\begin{equation}
\label{sRMT1}
d\mu_{\beta}(S^{(M)}) = \prod_{i<j} |{\rm e}^  {i\phi_i}-{\rm e}^ 
{i\phi_j}|^{\beta}
\prod_{i=1}^Md\phi_i d\Omega
\end{equation}
where $d\Omega$ is the solid angle on the $M$ dimensional unit hyper-sphere.

\begin{figure}
\begin{center}
\epsfxsize.5\textwidth%
\epsfbox{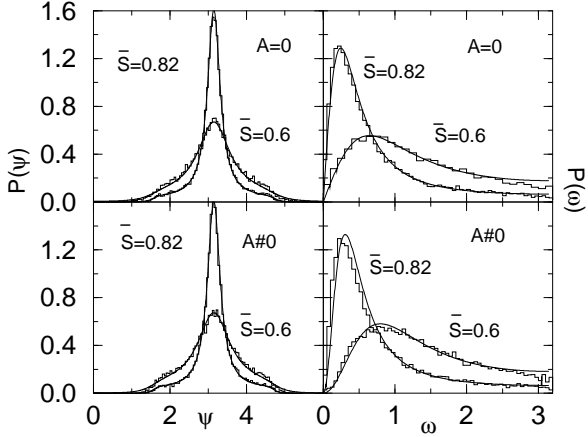}
\caption{ The ${\cal P}(\psi)$ and ${\cal P}(\omega)$ distributions for a $2
\times 2$ scattering matrix $S$. The solid lines correspond to the
predictions of
RMT (\ref{s2}). The upper panel correspond to $A=0$ while the lower panel to
$A\neq 0$.}
\end{center}
\end{figure}
For large $M$ values, a numerical check of this probability distribution is
prohibitive. However, for $M=2$ one can find easily the exact form of
(\ref{sRMT})
and compare with the numerical results. The resulting distribution of the
eigenphases $\phi_1,\phi_2$ of the $2\times 2$ scattering matrices of 
regular
graphs is given by
\begin{eqnarray}
\label{s2}
\hspace{-20mm}
& &d{\cal P}_{\bar S} (S^{(2)}) = \\
\hspace{-20mm}
&=& \frac {C_{\beta} \left(4v\over (v+1)^2\right)^{\beta+2}
\sin^{\beta}\left(\omega\over 2\right)}{\left[1-4{\bar 
S}\cos(\frac{\omega}2)
\cos\psi + 2{\bar S}^2 \left( 2\cos^2(\frac{\omega} 2)+\cos(2\psi)\right)-
4{\bar S}^3 \cos\psi \cos(\frac{\omega} 2)+{\bar 
S}^4\right]^{\frac{\beta+2}2}}
d\omega d\psi \nonumber
\end{eqnarray}
where we used the notation $\psi = \frac {\phi_1 +\phi_2}2$ and $\omega 
= \phi_1
- \phi_2$. Integrating Eq.~(\ref{s2}) over $\omega$, $\psi$ we get the 
corresponding
distribution functions ${\cal P}_{\bar S}(\omega)$ and ${\cal P}_{\bar 
S}(\psi)$.
Our numerical results for an ensample of $S$-matrices calculated for 
different
realizations of the lengths of the graphs are reported in figure~8 
together with
the RMT  predictions (\ref{s2}) for a regular graph with two channels. 
An overall
good agreement is seen both for $A=0$ and $A\neq 0$.

\subsection{\bf Partial delay times statistics}

The Wigner delay time  captures the time-dependent aspects of quantum 
scattering. It
can be interpreted  as the typical time an almost monochromatic wave 
packet remains
in the interaction  region. It is defined as
\begin{equation}
\label{wignersmith1}
T_W \ =\ {1\over 2 i M}{\rm tr} \left [  S^{(M)\,\,\dagger}
\frac{dS^{(M)}(k)}{dk} \right ]\ =
\ {1\over 2 M}
\sum_{i=1}^M \frac{\partial\phi_i(k)}{\partial k}\ ,
\end{equation}
where $\phi_i$ are the eigen-phases of the $S^{(M)}$-matrix.
The partial derivatives ${\partial \phi_i(k)\over \partial k }$ are the
{\it partial delay times} and their statistical properties were studied
extensively within the RMT \cite{FS96b,FS97,GM98}. For the one-channel 
case it
was found \cite{FS97,GM98} that the probability distribution of the scaled
(with the mean level spacing $\Delta$) partial delay times $T_i={\Delta 
\over
2\pi}{\partial \phi_i(k)\over \partial k }$ is
\begin{equation}
\label{wignersmith3}
{\cal P}_{\bar S}^{(\beta)}(T) =\frac{({\beta\over 2})^{\beta/2}}
{{\it \Gamma}({\beta\over 2})T^{2+\beta/2}}
\int_0 ^{2\pi} {\cal P}(\phi)^{1+\beta/2} {\rm e}^  {-{\beta\over 2T}{\cal
P}(\phi)}d\phi
\end{equation}
where ${\cal P}(\phi)$ is the Poisson Kernel Eq.~(\ref{sRMT}). The general
case of
$M$-equivalent open channels was studied in \cite{FS96b,FS97} where it was
found that
for broken time reversal symmetry the probability distribution of partial
delay
times is
\begin{equation}
\label{wignersmith4}
{\cal P}(T)= \frac{(-1)^M}{M! T^{M+2}}\frac {\partial^M}{\partial(T^{-1})^M}
\left[ {\rm e}^  {-g/T}I_0(T^{-1}{\sqrt(g^2-1)}) \right]
\end{equation}
where $I_0(x)$ stands for the modified Bessel function.

\begin{figure}
\begin{center}
\epsfxsize.5\textwidth%
\epsfbox{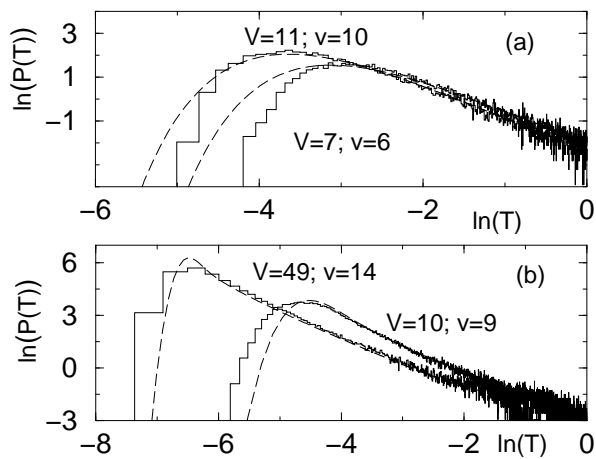}
\caption{
The distribution of the scaled partial delay times $T$ for various graphs with
Neumann boundary conditions. The dashed lines correspond to the RMT expectation
(\ref{wignersmith3},\ref{wignersmith4}): (a) One channel and $A=0$; (b) $M=V$
channels and $A\ne 0$.}
\end{center}
\end{figure}

To investigate the statistical properties of the partial delay times for our system we 
had calculated ${\cal P}(T)$ for various graphs. The resulting distributions are shown 
in figure~9  together with the RMT predictions (\ref{wignersmith3},\ref{wignersmith4}).
An overall agreement is evident.  Deviations appear at the short time regime (i.e. 
short  orbits), during which the ``chaotic" component due to multiple scattering is
not yet fully developed \cite {RBUS90}.


\subsection{\bf Conductance Distribution }

Due to the recent experimental investigation of electronic transport through mesoscopic
devices  \cite{MRWHG92}, the study  of the statistical properties of the conductance 
gained some importance. For a device connected to reservoirs by leads, the Landauer-
B\"uttiker formula relates its conductance $G$ to the transmission coefficient $T_G$ 
by the expression $G=(2{\rm e}^  2/\hbar) T_G$. When each lead supports one channel, 
the transmission coefficient can be written in terms of the $S^{(M)}-$matrix as
\begin{equation}
\label{landauer}
T_G= \sum_{i\neq j}^M |S_{i,j}^{(M)}|^2 = 1-|S_{j,j}^{(M)}|^2
\end{equation}
where $j$ is the input channel.
\begin{figure}
\begin{center}
\epsfxsize.5\textwidth%
\epsfbox{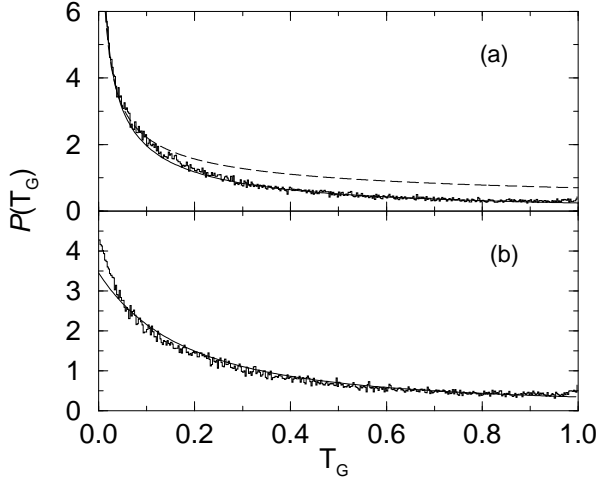}
\caption{ Conductance distribution ${\cal P}(T_G)$ for a graph with $M=2$
channels: (a) Time reversal symmetry is preserved ($A=0)$. The solid line is
the RMT results Eq.~(\ref{ptrans}) where we had used Eq.~(\ref{s2}) for the
Poisson Kernel while the dashed line is the approximate expression (\ref{Slarge1});
(b) Broken time reversal symmetry ($A\ne 0$). The solid line is the RMT result
Eq.~(\ref{Slarge2}).}
\end{center}
\end{figure}

In the absence of direct processes, the distribution of conductance ${\cal P}(T_G)$ 
for arbitrarily number of channels was worked out within RMT, and the results 
describe in a satisfactory way both the numerical calculations and the experimental 
data (for a review see \cite{MB99} and references therein). However in cases where 
direct processes appear significantly, one must use the Poisson's kernel (\ref{sRMT}) 
in its full  generality. This is exactly the case with Neumann graphs since ${\bar S} 
\neq 0$. The  probability distribution of the transmission $T_G$ is then
\begin{equation}
\label{ptrans}
{\cal P} (T_G) = \int \delta(T_G-\sum_{i\neq j} |S_{i,j}|^2) {\cal P}_{\bar
S}(S)
d\mu(S).
\end{equation}

For the case $M=2$ and for a diagonal ${\bar S}-$matrix (only direct
reflection) with
equivalent channels ${\bar S}_{11}={\bar S}_{22}={\bar S}$,
Eq.~(\ref{ptrans}) can be
worked out analytically \cite{BB94} in the limit of strong reflection and
$T_G\ll 1$.
The resulting expression is:
\begin{equation}
\label{Slarge1}
{\cal P}_{\beta=1}(T_G)=
{8\over \pi^2(1-{\bar S}^2)} T_G^{-1/2}\quad , \quad  T_G\ll (1-{\bar 
S}^2)^2
\end{equation}
For $\beta=2$ one can  compute in a close form the whole distribution
\cite{MB99}
\begin{equation}
\label{Slarge2}
{\cal P}_{\beta=2}(T_G)= (1-{\bar S}^2) \frac{(1-{\bar S}^4)^2+2{\bar 
S}^2 (1+
{\bar S}^4)T_G+ 4{\bar S}^4T_G^2}{((1-{\bar S}^2)^2 + 4{\bar S}^2 
T_G)^{5/2}}.
\end{equation}

Our numerical results for a regular graph with two leads are plotted in 
figure~10
together with the RMT results 
(\ref{ptrans},\ref{Slarge1},\ref{Slarge2}). Notice
that,  the small conductances are emphasized, because of the presence of 
direct
reflection  and no direct transmission. At the same time, the most 
pronounced
differences between  the two symmetry classes are for $T_G\ll 1$, where for
$\beta=1$ the conductance distribution diverges while it take a finite 
value for
$\beta = 2$.


\section{\bf Scattering from star-graphs.}
\label{sec:star}

So far, we have studied scattering from well connected graphs, and we have shown that
many statistical properties of the scattering matrix are described by RMT. We shall
dedicate this section to demonstrate the statistical properties of the $S-$matrix for
graphs which have non-uniform connectivity.

A representative example of this category are the ``star" (or ``Hydra") graphs 
\cite{KS99,BK99}. They consist of $v_0$ bonds, all of which emanate from a single 
common vertex labeled with the index $i=0$. The vertex at $i=0$, will be referred
to in the sequel as the  {\it head}. The total number of vertices for such a graph 
is $V=v_0+1$, and the vertices at the end of the bonds will be labeled by $i=1
\cdots v_0$. The star is a bipartite graph, a property which implies e.g., that there
exists no periodic orbits of odd period \cite{KS99}. To turn a star graph  into a 
scattering system we add a lead to its head.

It is a simple matter to derive the scattering matrix $S=S^{(M=1)}$ for a Neumann star. 
It reduces to a phase factor, 
\begin{equation}
\label{contH}
S(k)\equiv {\rm e}^  {i\phi(k)} = {-\sum_{i=1}^{v_0} \tan(kL_i) +
i \over \sum_{i=1}^{v_0} \tan(kL_i) + i}
\end{equation}

The spectrum $\{k_n\}$ of the close system can be identified as the set of wave-vectors
for which $S(k)$ equals  1, which implies that no current flows in the lead. The 
resulting quantization condition is
\begin{equation}
\label{eq:starb}
1-S(k)=0\longleftrightarrow {2\sum_{i=1}^{v_0} \tan(k_nL_i) \over
\sum_{i=1}^{v_0}
\tan(kL_i) + i}=0\quad,
\end{equation}
which is satisfied once $\sum_{i=1}^{v_0} \tan(k_nL_i)=0$. This is 
identical with
the condition derived in \cite{KS99}.

The poles $\{\kappa_n\}$ are the complex zeros of
\begin{equation}
\label{openH}
\sum_{i=1}^{v_0}\tan(\kappa_nL_i) + i =0.
\end{equation}
To first order in $\frac {1}{v_0}$, we get
\begin{equation}
\label{pertH}
\Gamma_n^{(1)} = {1\over\sum_{i=1}^{v_0}{L_i\over \cos(2k_nL_i) +1}}\
\end{equation}
which can be used as a starting point for the exact evaluation of the 
poles. For the
latter one has to perform a self consistent search for the complex zeros 
of the
secular equation (\ref{openH}). This is a time consuming process and the 
correct
choice of the initial conditions is very important.

In figure~11  we present our numerical results for the distribution of 
rescaled
resonance widths ${\cal P}(\gamma)$ for a star with $v_0=20$. The data 
are in
excellent agreement with the RMT expectation given in Eq.~(\ref{chi2}). 
We point
that in this case the coupling  to the continuum is weak since $g\simeq 
10 \gg 1$
and therefore the $\chi^2$-distribution  with $M=1$ is applicable.

\begin{figure}
\begin{center}
\epsfxsize.5\textwidth%
\epsfbox{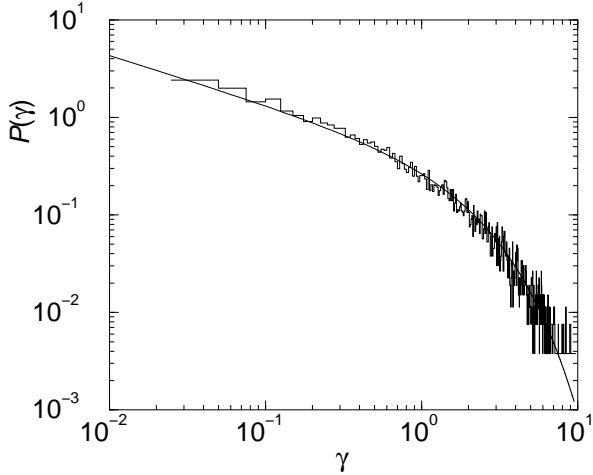}
\caption{ The rescaled resonance width distribution ${\cal P}(\gamma)$ for
a star graph
with $v_0=20$. The solid line is the RMT prediction Eq.~(\ref{chi2}).}
\end{center}
\end{figure}

Using Eq.~(\ref{contH}) we get the following relation for the scaled 
delay times
\begin{equation}
\label{delayH}
T(k) ={\Delta\over 2\pi}{2\sum_{i=1}^{v_0} {L_i\over \cos^2(kL_i)}\over 1+
\left(\sum_{i=1}^{v_0}\tan k L_i\right)^2}
\end{equation}
which can be used to generate ${\cal P}(T)$. The latter is reported in figure~12
together with the RMT prediction ~(\ref{wignersmith4}). We notice that although
the tail of the distribution agrees reasonably well with the RMT prediction, there
are considerable deviations at the origin.

A peculiarity of the star graph is that the mean time delay is twice larger than the
expected one from semiclassical considerations. To be more specific, for a generic
chaotic system coupled to $M$ channels, one has \cite{DS92a}
\begin{equation}
\label{taumH}
\langle T\rangle _k \simeq {M+1\over M} {\Delta\over 2\pi} \langle l_p\rangle _k
\end{equation}
where $\langle l_p\rangle _k\simeq \Gamma_{cl}^{-1}$ is the average length of the 
classical paths inside the interaction area. Thus for the star with $M=1$, we would 
expect based on Eq.~(\ref{taumH}) that $\langle T\rangle _k \simeq 2 \Gamma_{cl}$. 
However, due to the fact that all the periodic orbits on the  star graph are self 
tracing we get an additional factor of $2$ and thus
\begin{equation}
\label{mtHc}
\langle T\rangle _k= {\Delta\over 2\pi} 4 \Gamma_{cl}^{-1}.
\end{equation}

\begin{figure}
\begin{center}
\epsfxsize.5\textwidth%
\epsfbox{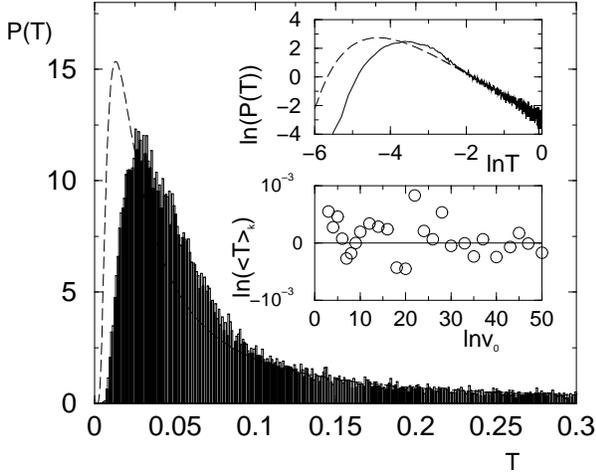}
\caption {The distribution of the scaled partial delay times ${\cal P}(T)$ for a star 
graph with $v_0=20$. The dashed line is the RMT prediction Eq.~(\ref{wignersmith3}).
In the upper inset we show the same data in a double logarithmic plot. In the lower 
inset we plot the numerical results for $<T>_k$ ($\circ$). The solid line is the 
asymptotic value $1$ expected from semiclassical considerations.}
\end{center}
\end{figure}

The corresponding classical decay rate $\Gamma_{cl}$ can be found exactly by a direct
evaluation of the eigenvalues of the classical evolution operator ${\tilde U}$. One 
can easily show that
\begin{equation}
\label{UH1}
{\tilde U} =\left( \begin{array}{ll}
  0 & {\bf 1}  \\
({\tilde \sigma^{(0)}}) ^2 & 0
\end{array}  \right)
\end{equation}
where ${\tilde \sigma}^{(0)}$ is the $v_0\times v_0$ vertex scattering matrix at
the star head as defined in Eq.~(\ref{Neumann}), while ${\bf 1}$ denotes the
$v_0\times v_0$ identity matrix. The square of the classical evolution operator
${\tilde U}^2$ has a block diagonal form
\begin{equation}
\label{UH2}
{\tilde U}^2 =\left( \begin{array}{ll}
({\tilde \sigma^{(0)}}) ^2 & 0 \\
0 & ({\tilde \sigma^{(0)}}) ^2
\end{array}  \right).
\end{equation}
and its spectrum consists of the values $1-{4\over (1+v_0)^2}$ with multiplicity
2 and $1-{4\over (1+v_0)}$ with multiplicity $2v_0-2$. Therefore the spectrum of
${\tilde U}$ is
\begin{eqnarray}
\label{UH3}
\lambda _u &=& \pm \sqrt {1-{4\over(1+v_0)^2}}\\
            &=& \pm \sqrt {1-{4\over 1+v_0}} \quad \quad {\rm with}\,\, {\rm
multiplicity}
\,\, v_0-1\nonumber.
\end{eqnarray}
For short times where the classical evolution is applicable, the dominant eigenvalue
is $\lambda_u=\sqrt{1-{4\over 1+v_0}}$ leading to a classical decay rate 
\begin{equation}
\label{Hgammac}
\Gamma_{cl} \simeq {2\over 1+v_0}.
\end{equation}
Substituting Eq.~(\ref{Hgammac}) we get eventually that $\langle T\rangle _k = {1+
v_0\over v_0} \ \overrightarrow{\scriptstyle v_0{\rightarrow \infty }}\ 1$ which 
indicate that the mean time a particle spend inside the interaction regime is
proportional to the Heisenberg time. Our numerical calculations reported in the lower 
inset of Fig.~12 agrees nicely with the semiclassical prediction ~(\ref{mtHc}).

Finally, we analyze the distribution of $S$ when generated over different realizations 
of the lengths of the bond. For the one channel case this is equivalent with the 
distribution ${\cal P} (\phi)$ of the phase of the $S$-matrix. To derive the latter, 
it is convenient to rewrite the $S$-matrix in the following form
\begin{equation}
\label{hydraPs}
S\equiv \exp(i\phi) =(-K + i)/(K + i),
\end{equation}
with $ K = \sum_{i=1}^v \tan(kL_i) $. The probability distribution of
$K$ is
\begin{eqnarray}
\label{hydra2}
{\cal P}(K)& = &\left \langle \delta\left(K-\sum_{i=1}^{v}
\tan (kL_i) \right)\right \rangle_{L}\nonumber \\
& = & \frac 1{2\pi} \int {\rm e}^  {iKx} dx \left( \frac1 {\Delta L}
\int_{L_{min}}^{L_{max}} dL {\rm e}^  {-ix\tan(kL)}\right)^v \nonumber \\
&=& \frac 1{\pi}
\frac {v}{v^2+K^2} \ .
\end{eqnarray}
Thus, with $\bar {S} \equiv \langle S\rangle =\frac{1-v}{1+v}$, we get
\begin{equation}
\label{hydra3}
{\cal P}(\phi) = {\cal P}(K) \left|\frac{dK}{d\phi}\right| =
\frac 1{2\pi} \frac{1-{\bar S}^2}{1+{\bar S}^2-2{\bar S} \cos\phi} \ .
\end{equation}
Equation ~(\ref{hydra3}) reproduces the Poisson's kernel for a one-channel
scattering
matrix, derived in the framework of RMT \cite{FS97}. The conditions under
which this
result is derived in \cite{FS97} are fulfilled {\it exactly} in the present
case. Our numerical results are reported in figure~13 and are in
excellent agreement  with the theoretical prediction Eq.~(\ref{hydra3}).

\begin{figure}
\begin{center}
\epsfxsize.5\textwidth%
\epsfbox{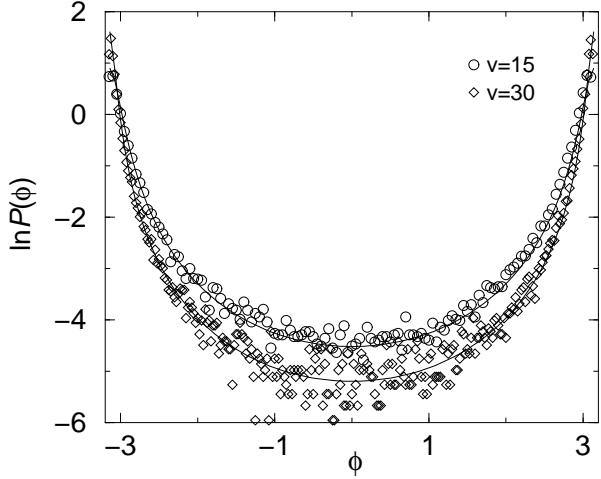}
\end{center}
\label{fig:starphase}
\caption{The distribution of phases ${\cal P}(\phi)$ of the $S_H$-matrix
for two stars with $v_0=15$ and $v_0=30$. The solid lines are the 
corresponding
theoretical predictions (\ref{hydra3}).}
\end{figure}


\section{\bf Conclusions}
\label{sec:conclusions}

In this paper, we turned quantum graphs into scattering systems. We show that they
combine the desirable features of both behaving ``typically" and being mathematically
simple. Thus, we propose them as a convenient tool to study generic behavior of
chaotic scattering systems.

The classical dynamics on an open graph was defined, and the classical staying
probability was shown to decay in an exponential way. The resulting classical
escape rate was calculated and used to describe the properties of the corresponding
quantum system. The scattering matrix was written in terms of classical orbits and
an exact trace formula for the resonance density was found. A gap for the resonance
widths has been obtained for ``generic" graphs and its absence was explained for
Neumann graphs. An analysis of the cross section autocorrelation function was 
performed and its non-universal characteristics were explained in terms of the short
classical scattering trajectories. Finally, due to the relative ease by which a large
number of numerical data can be computed for the graph models, we had performed a
detail statistical analysis of delay times, resonance widths and distribution of the
$S$-matrix. Our results compares well with the predictions of RMT indicating that our
model can be used in order to understand the origin of the connection between RMT and
the underlying classical chaotic dynamics.

The results reported here, complete our previous investigations on graphs. We conclude
that quantum graphs may serve as a convenient paradigm in the area of quantum chaos,
both for spectral and scattering studies.

\section{\bf Acknowledgments}

We acknowledge many useful discussions with Y. Fyodorov and H. Schanz. This work was
supported by the Minerva Center for Nonlinear Physics, and an Israel Science Foundation
Grant. (T.~K.) acknowledges a postdoctoral fellowship from the Feinberg School, The 
Weizmann Institute of Science.


\end{document}